# Realisation of a Frustrated 3D Magnetic Nanowire Lattice

Andrew May, Matthew Hunt, Arjen Van Den Berg, Alaa Hejazi and Sam Ladak[*]

School of Physics and Astronomy, Cardiff University, Cardiff, U.K. CF24 3AA

* Email: LadakS@cardiff.ac.uk

**Patterning nanomagnets in three-dimensions presents a new paradigm in condensed matter physics and allows access to a plethora of fundamental phenomena including robust spin textures, magnetic metamaterials that are home to defects carrying magnetic charge and ultrahigh density devices that store information in three-dimensions. However, the nanostructuring of functional magnetic materials into complex three-dimensional geometries has thus far proven to be a formidable challenge. Here we show magnetic nanowires can be arranged into 3D frustrated magnetic nanowire lattices by using a combination of 3D polymer nanoprinting and metallic deposition. The fabricated nanowires are single domain and they switch via nucleation and propagation of domain walls. Deep nanoscale magnetic imaging and finite element simulations elucidate the spin texture present on the 3D nanostructured lattice. Our study demonstrates a generic platform for the production of 3D nanostructured magnetic materials allowing the realisation of racetrack memory devices and 3D nanostructured systems that mimic bulk frustrated crystals.**

Frustration is defined as the inability for all pairwise interactions to be simultaneously satisfied [1] and is a phenomenon that impacts a wide variety of areas. For example, frustration is important in the production of solar flares [2], in the folding of biological molecules [3]and in the bonding of water ice [4]. Realising model systems for studying frustration will help to elucidate a range of complex phenomena in science. One such model system is the spin-ice materials which have a pyrochlore structure with Ising spins located on corner-sharing tetrahedra [5]. The local crystal field forces the spins to point along the local [111] direction and frustration leads to a local ordering principle known as the ice-rule [5], whereby two spins point into a tetrahedron and two spins point out. This leads to an extensive ground state degeneracy and an associated residual entropy found at low temperature. In addition, spin-ice systems have been found to harbour monopole-like defects that propagate through the material and interact via a magnetic Coulomb's law [6, 7]. In 2006 it was shown that arranging magnetic nanowires into frustrated geometries allows much of the physics of spin-ice to be captured [8]. In order to emulate the geometry of spin-ice, magnetic nanowires were arranged in a square lattice with four spins meeting at a vertex. Since magnetic nanowires are essentially bipolar, they are considered effective Ising spins [8]. The advantage of these "artificial spin-ice" (ASI) systems is that the magnetic islands have dimensions of order 100nm and thus can be probed with magnetic microscopy. This has led to a wealth of studies [9] probing the physics of ordering [8, 10-15], thermodynamics [16, 17] and the nature of monopole-like defects in these systems [18-25]. However, it was acknowledged early on that square ASI is not a perfect analogy with the bulk counterpart because the energy of interaction between neighbouring islands depends upon those chosen [26]. Layered ASI systems have partially solved this problem [17, 27, 28]. By offsetting one of the sublattices by some height such that the energy of interaction between all nearest neighbours becomes equivalent, a magnetic Coulomb phase has been realised [29]. Unfortunately, such layered systems still do not capture the exact physics of bulk systems. This is due to the magnetic islands not being arranged in the exact geometry of the spins in the bulk material, along the local [111] direction[18]. Ideally one would like to produce a diamond-bond 3D nanostructured magnetic lattice which maps perfectly onto the spin-ice system.

Quite separately, there has been a recent surge in the interest upon 3D nanostructured magnetic materials [30]. This has partially been driven by theoretical studies that show 3D nanostructuring of magnetic materials [30] offers a new means to control the arrangement of spins in ordered magnetic materials by harnessing geometry dependent magnetic energies and those provided by more subtle curvature-driven effects [31]. In addition, applications-driven research remains, with the realisation that 3D nanostructuring of magnetic materials will directly enable new racetrack-type memories [32].

Thus far there have been three main ways to realise 3D magnetic nanostructures. The simplest involves electrodeposition of a magnetic material into a porous template and this usually yields pure nanowires of cylindrical geometry [33-35]. Here, utilising electrodeposition allows a large variety of magnetic materials to be used in filling the templates providing enormous flexibility. A disadvantage of this approach is that the templates usually have cylindrical pores, limiting the exploration of more complex 3D geometries. Focussed electron beam deposition (FEBD)[36] is a newer methodology that harnesses the interaction between an electron beam and a chemical precursor. The reaction takes place at the point of electron focus and by moving this with respect to the substrate, 3D nanostructures can be fabricated. The method is powerful since it allows the realisation of almost any 3D geometry, potentially allowing one to probe the rich physics in 3D magnetic nanostructured systems. However, when the beam interacts with the chemical precursor and any background gases within the chamber, it also leads to the deposition of nonmagnetic contaminants such as carbon and oxygen. These become embedded within the nanowire and impact the magnetic properties of the system. Despite this a number of ground-breaking studies have explored the use of FEBD in the realisation of 3D magnetic nanostructures. In a pioneering study, Co nanowires in 3D geometries were produced and measured using magneto-optical Kerr effect (MOKE) and magnetic force microscopy [37]. Unfortunately, the nanowires were found to be only 85% Co and the nanowires exhibited irreproducible switching characteristics. More recent studies have focussed upon the growth of more complex 3D nano-cube and nano-tree geometries [38, 39]. Here, detailed micro-Hall measurements were used to measure the magnetic switching within the structures [38, 39]. Unfortunately, the structures were found to be 64% metal (Co and Fe) and measurements indicated the individual nanowires were not behaving single domain [38]. A further study by the same authors also concluded the magnetisation within nanowires was inhomogeneous with curling magnetisation states [39]. Recent work has shown that focussed electron beam deposition of Pt when coupled with NiFe evaporation can yield planar magnetic nanowires in 3D geometries, which switch via domain wall motion [40]. The magnetic switching within these simple systems was measured using a novel dark-field magneto-optical Kerr effect magnetometry technique. This methodology is very exciting but has not been applied to the manufacture of complex 3D nanostructured lattices. Finally, two-photon lithography and electrodeposition has been used to fabricate complex 3D magnetic nanostructures [41, 42] which were also very pure (>95% Co). Unfortunately, the structures made within this study were relatively large (>400nm) and shown to be multi-domain [41].

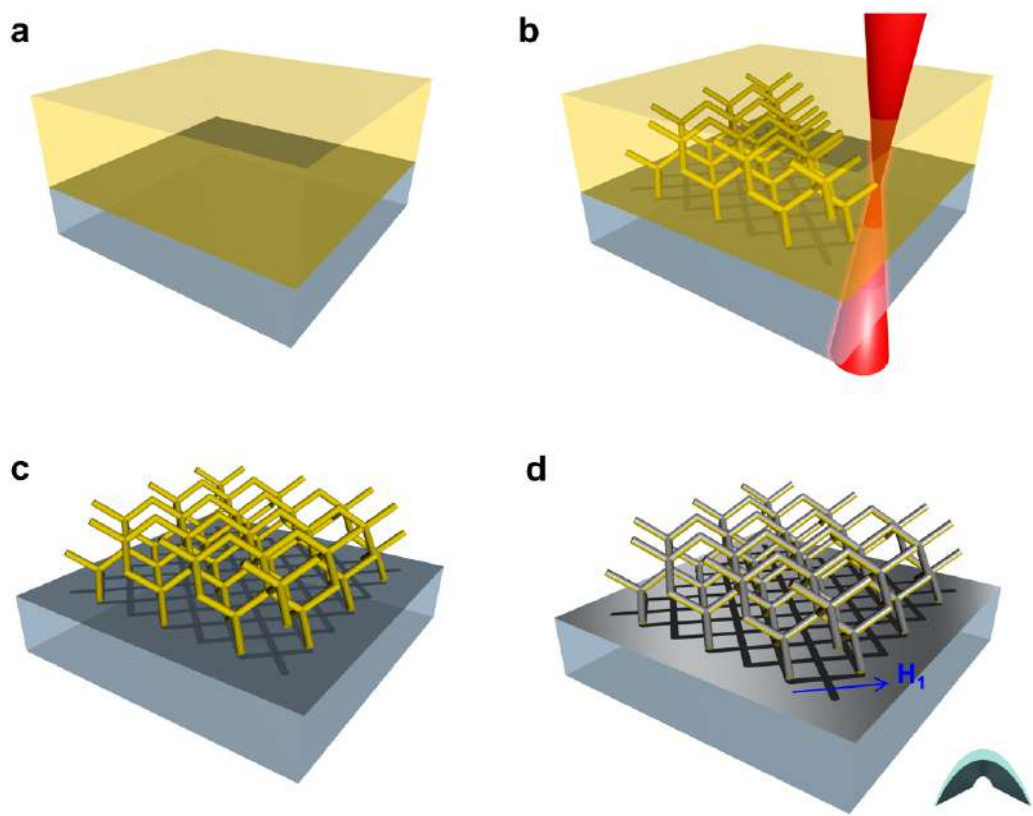

**Figure 1: Schematic of fabrication methodology. (a) Negative-tone resist is drop cast onto the substrate. (b) Two-photon lithography is used to define a diamond bond lattice within the resist. (c) Development removes unexposed resist revealing the freestanding diamond lattice structure. (d) Magnetic material is evaporated onto the lattice structure. The exposed surface of the lattice is coated yielding a 3D network of magnetic nanowires. Inset: The nanowires have a crescent-shaped geometry as shown.**

Measurement of the 3D spin-texture within magnetic materials is also a significant challenge. A recent ground breaking measurement demonstrated the use of hard x-ray tomography to produce a 3D reconstruction of the spin-texture within a micron sized $GdCo_2$ cylinder [43]. Combining 3D lithography techniques with nanoscale magnetic imaging offers the possibility of directly visualising ground state ordering in complex 3D artificial crystals. In particular, replicating the exact spin arrangement seen in the bulk frustrated materials known as spin-ice [5] will allow realisation of new model systems for studying frustration.

Here we demonstrate a new methodology to realise single domain magnetic nanowires within complex frustrated 3D nanostructured lattices (3DNL). With these newly realised samples, we directly observe via optical magnetometry and finite element simulations, switching by domain wall motion through the 3DNL. In addition, simulations and nanoscale magnetic imaging allows the determination of the exact spin texture upon the 3DNL. Our results suggest these systems may be considered a 3D ASI system.

 Results

The fabrication methodology is sketched in Fig 1. Two-photon lithography (TPL) is used to manufacture the desired 3D geometric network. Any geometry can be chosen so long as the upper surface of the desired nanowires is not masked. For this study, a diamond-bond lattice geometry has been chosen to demonstrate that complex nanowire networks can be created and also to capture the geometric arrangement of spins within bulk frustrated materials known as spin-ice [5]. After the polymer network has been created, evaporation is used to deposit a magnetic material (Ni$_{81}$Fe$_{19}$) upon the nanowires. Evaporation onto a curved geometry yields magnetic nanowires of crescent-shaped cross-section [44, 45] (Fig 1d inset). The nanowire network of interest is 10 μm above the substrate minimising the stray field from any magnetic material on the planar substrate.

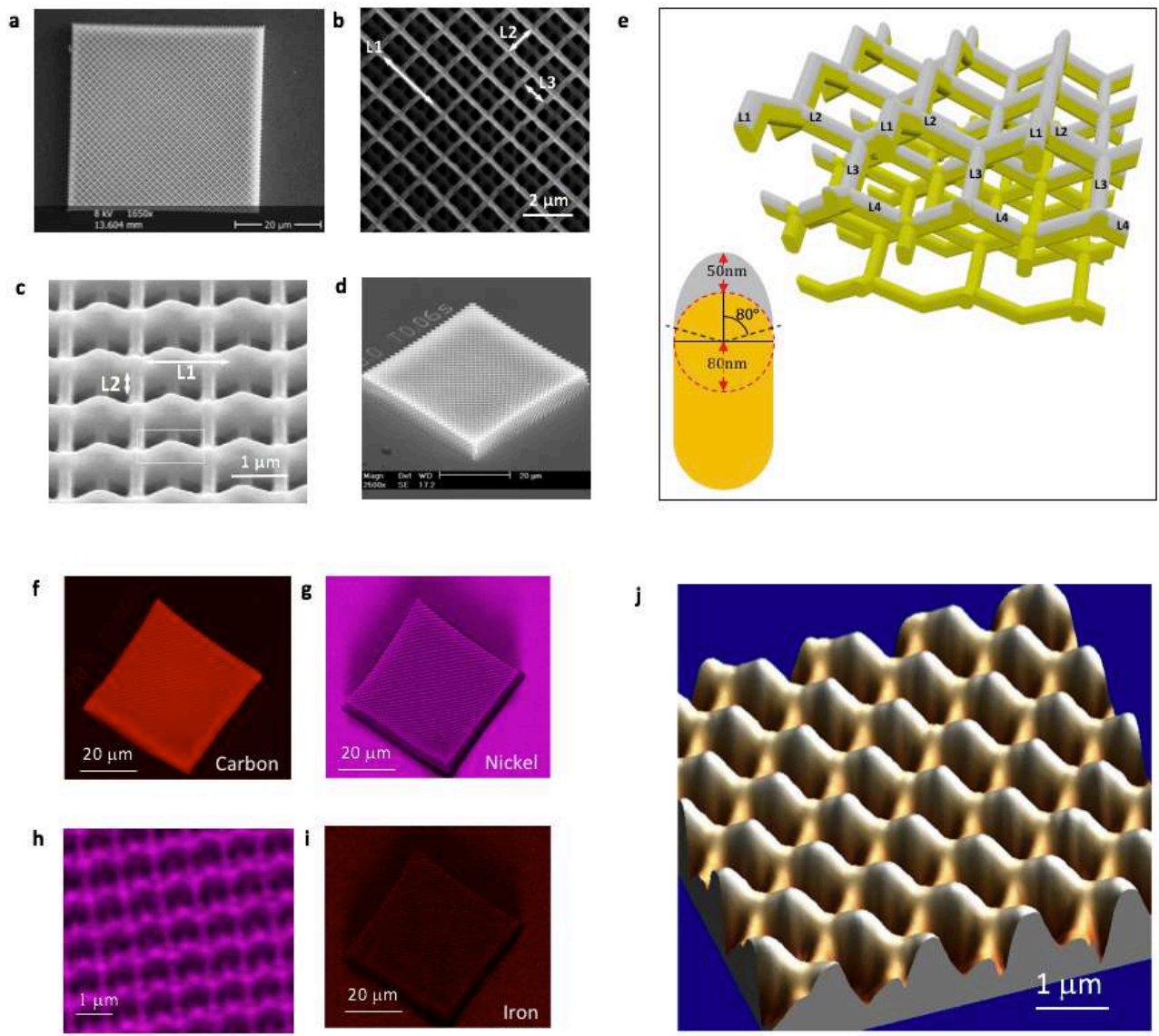

**Figure 2: Physical and material properties of the 3D magnetic nanowire lattice. (a) Scanning electron microscope image of the entire array (Top view). (b) Close-up image of the nanowire lattice. L1, L2 and L3 indicate the different sub-lattice layers. (c-d) Angled images of the 3D nanowire lattice. The dotted box in c indicates a single bipod unit. (e) A schematic of the realised 3D magnetic nanowire lattice. Inset: Cross-sectional geometry of the nanowires. Yellow depicts polymer and grey depicts $Ni_{81}Fe_{19}$. (f) Energy dispersive x-ray spectroscopy images for carbon, (g) for nickel, (h) for nickel at higher magnification and (i) for iron. (j) Atomic force microscopy image of the nanowire lattice. Scale preserved in both directions.**

**Physical characterisation.** The 3DNL arrays have dimensions 50μm × 50μm × 10μm (Fig 2a). The Ni$_{81}$Fe$_{19}$ nanowires have an inner radius of curvature of 80nm yielding an effective outer arc length of 330nm and have length of approximately 1000nm. Since evaporation is used to deposit the magnetic material, the resulting magnetic nanowires are continuous for three sub-lattice layers (L1, L2, L3) into the z-direction (Fig 2b). Fig 2c/2d show angled views of the nanowire network. Both images show the 3D geometry of the network. In Fig 2c it is clear that the polymer nanowire geometry deviates from the perfect cylindrical geometry proposed in Fig 1 and the nanowires are found to have a greater axial size than lateral size. This effect is well known in two-photon lithography and it is due to the ellipsoidal point spread function of the laser at focus [41]. Some methodologies allow this to be optimised (see discussion) but in this study, since evaporation simply coats the top surface of the polymer nanowire, it is not a concern. A schematic of the grown 3DNL is shown in Fig 2e, with an individual wire cross-section, based upon SEM data shown in the inset. The Ni$_{81}$Fe$_{19}$ is seen to subtend an arc of approximately 80° either side of the centre, beyond which the polymer has vertical sidewalls. In Fig 2e the continuity of the L1, L2 and L3 layers can be seen within the lattice. The fourth layer (L4) is seen to have a discontinuity at the lower junction as a direct result of shadowing from L1. Below L4 the polymer nanowires are perfectly shadowed by layers above, to within the positional reproducibility of the galvanometer (~20nm). From geometric considerations, any magnetic material present on L5 or lower is expected to have a magnetic moment of less than 10$^{-16}$ Am$^2$. Accordingly, we find that the dipolar energy of interaction between L5 and L4 wires is two orders of magnitude lower than the energy of interaction between wires on upper layers. Thus, we conclude that any dipolar interaction between potential material upon L5 or lower, and nanowires L1, L2, L3 and L4, is negligible. Fig 2e also shows the interesting geometry of the lattice at junctions (for example between L1 and L2). Due to the extended axial nanowire dimension, a constriction is expected at the junction which may impact domain wall propagation within the lattice.

Energy dispersive x-ray spectroscopy (EDX) allows one to visualise the chemical constituents upon the 3DNL. The main constituent of the resist is carbon and mapping this element (Fig 2f) yields an image of the underlying polymer template. The slight deformation of the array edge is due to long exposure to an electron beam (~8 hrs), which was required to obtain sufficient signal. Mapping the Ni content across the array (Fig 2g, 2h) shows that the deposited magnetic material is taking the 3D form of the polymer template and that the magnetic nanowire network is continuous. Finally, obtaining the Fe signal (Fig 2i) allows the composition of the magnetic nanowires to be determined. The ratio of nickel to iron using EDX analysis is found to be 4.26, in excellent agreement with the composition of $Ni_{81}Fe_{19}$. Careful optimisation of feedback parameters also allows direct atomic force microscopy of the 3DNL surface (Fig 2j). Both the upper L1 and lower L2 nanowire layers are resolved.

A key question with such lattices is the extent to which nanowire quality (eg. geometry, composition and roughness) varies with sub-lattice layers. The polymer nanowires were written utilising a power compensation routine to ensure minimal difference between exposure for different layers. SEM analysis shows mean nanowire widths of 194±11nm, 191±11nm and 194±12nm for L1 (highest), L2 and L3 (lowest) respectively. This indicates that the power compensation routine is working correctly and minimal variation in width is observed between layers. Surface roughness of wires is more difficult to measure for all layers. Visual inspection of each layer with SEM does not show any qualitative variation. Atomic force microscopy images were used to carry out roughness analysis upon L1 and L2 nanowires, yielding values of 10.8±4.3nm and 16.1±3.2nm respectively. This suggests that the nanowire roughness across different layers does not vary significantly within error.

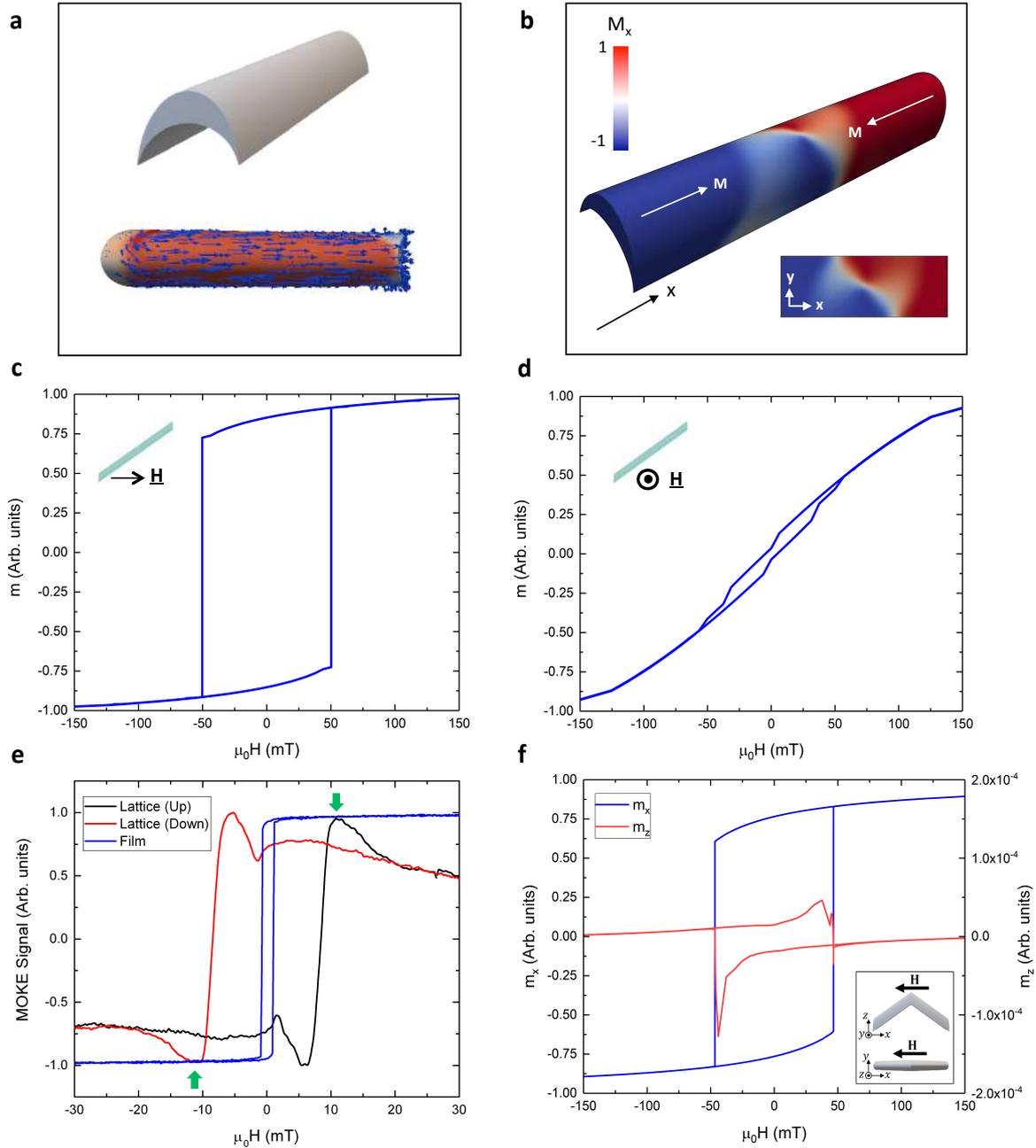

**Figure 3:** Finite element simulations and experimental magnetometry upon the 3D magnetic nanowire lattice. (a) Top: The geometry of the magnetic nanowires. Bottom: Simulated remanent magnetisation profile after relaxation. (b) Simulated relaxation with opposing magnetisation showing a vortex domain wall. (c) Simulated hysteresis loop of single magnetic nanowire with field applied at 35° to the wire long axis. An abrupt switching transition is seen, indicative of domain wall motion. (d) Simulated loop with field applied perpendicular to long axis projection. (e) Longitudinal magneto-optical Kerr effect loop of the thin-film upon the substrate (blue) and the 3D nanowire lattice (black: up sweep, red: down sweep). The nanowire lattice is found to have a coercivity one order of magnitude higher than the film as expected for a nanostructured magnetic system. Green arrows indicate the polar magneto-optical Kerr effect signal. (f) Simulated loop for bipod structure with field applied as shown in inset.

**Finite-Element Simulations and Optical Magnetometry.** Finite element simulations were performed in order to investigate the switching within the underlying nanowires. For details of geometry construction see Extended Data Fig 2. We start by allowing the nanowire magnetisation to relax in the absence of an applied magnetic field. The result is a single domain state, with magnetisation lying along the long axis of the wire (Fig 3a/bottom). Planar magnetic nanowires are well known to switch via domain wall motion [46] and the domain wall type depends upon wire width and thickness [47]. The nanowires that make up the 3DNL are distinct from their planar counterparts since they have both a graded thickness and cross-sectional curvature. A simulation was performed in order to investigate the domain wall type seen within crescent-shaped nanowires. A vortex domain wall relaxes in the region between two opposing magnetisation directions (Fig 3b). Figure 3c shows a simulated hysteresis loop of the nanowire when the field is applied at an angle of 35° with respect to the long axis. This field geometry matches the angle of the field with respect to upper wires in optical magnetometry experiments. The loop exhibits an abrupt transition, indicative of domain wall motion, at a magnetic field of approximately 50mT. Rotating the magnetic field by 90° such that it has no component along the wire long-axis, yields a hard-axis type loop with low remanence (Fig 3d), demonstrating the high shape anisotropy exhibited by the wire.

Measuring the magnetic properties of 3D magnetic nanostructures is a key challenge. Here, focussed MOKE magnetometry is utilised to measure the switching of the 3DNL. A distinct advantage of realising dense 3DNL samples is the optical scattering centres provided by buried wires. Such a system significantly reduces the specular reflection from the substrate. In order to probe the in-plane magnetisation of the 3DNL, MOKE measurements were performed with the wave vector at 45° angle of incidence and with the field ($H_1$) in the substrate plane (Fig 3e red/black curve). After careful alignment of the laser spot, we measure the 3DNL with only minimal signal from the substrate. The loop exhibits a sharp transition indicative of domain wall motion, an enhanced coercive field of approximately 8.2mT, indicative of nanostructuring

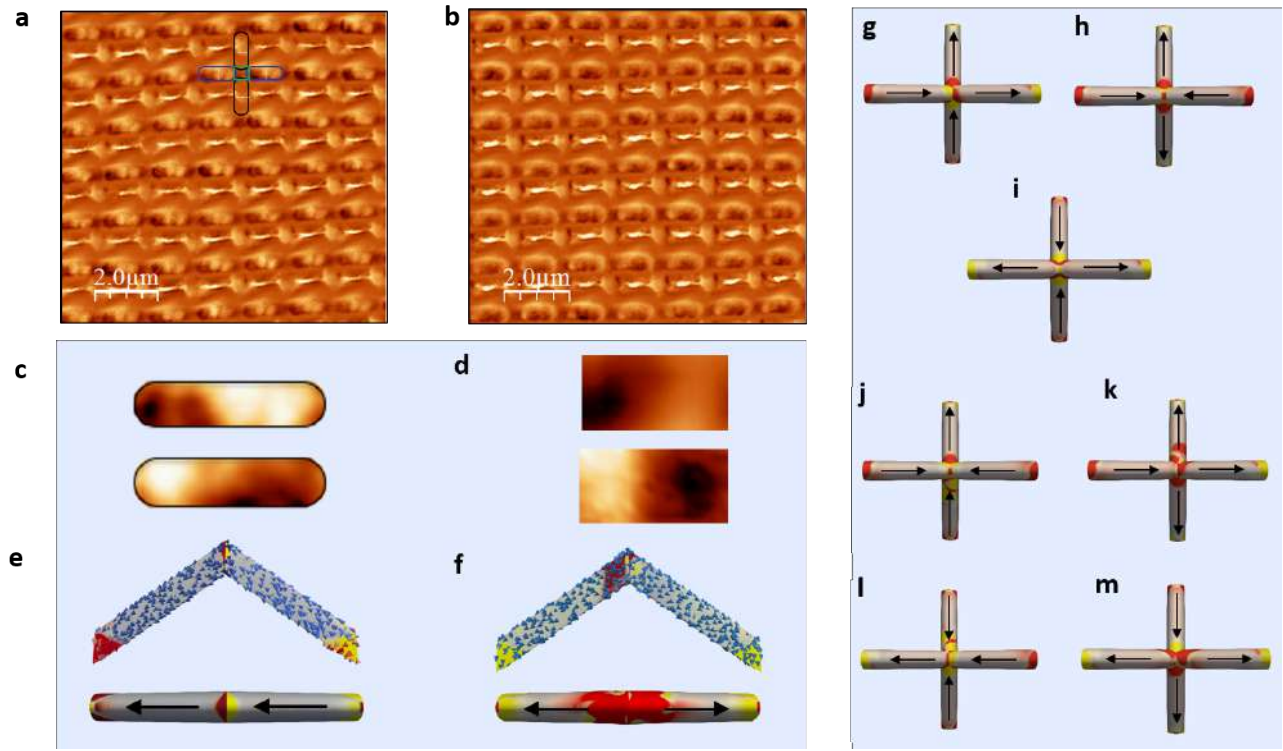

**Figure 4: Determining the magnetic configuration upon the 3D nanostructured lattice.** (a) Magnetic force microscopy image of the 3D nanolattice surface after field has been applied along projection of upper wire long axis ($H_1$). (b) Image of same area but with reversed tip magnetisation. (c) Top: Magnetic contrast of an individual bipod unit upon the surface. Bottom Magnetic contrast of bipod unit after tip magnetisation has been reversed. (d) Top: Magnetic contrast taken at vertex, where four wires meet. Bottom: Magnetic contrast taken at vertex after tip magnetisation has been reversed. (e-f) Finite element simulations of bipod structures whereby the wires were allowed to relax with magnetisation in head-to-tail arrangement and tail-to-tail arrangement respectively. (g-i) Finite element simulations showing low energy 2-in/2-out states that can be realised within a magnetic tetrapod structure composed of single domain wires. (j-m) Finite element simulations showing higher energy 3-in/1-out and 3-out/1-in states. Red and yellow indicate MFM contrast $\rho = \nabla \cdot \mathbf{M}$

and a high remanence. The loop also exhibits peaks close to the coercive field (Green arrow, Fig 3e) due to the polar MOKE effect (See Supplementary information). Moving the laser spot off the 3DNL and onto the sheet film produces a loop characteristic of bulk $Ni_{81}Fe_{19}$ (Fig 3e Blue). A key question with respect to the MOKE measurement upon the 3DNL is the extent to which it is probing each of the individual layers. In order to answer this question, the experimental data was compared to simulated bipod and tetrapod loops. A tetrapod is a unit cell for combined L1 and L2 layers and its simulated loop (See Extended Data Fig 3) has an extended tail leading to saturation as well as a low remanence (<0.5$M_s$), not seen in our experimental data. The simulated bipod loop (Fig 3f) has both longitudinal ($M_x$) and polar ($M_z$) contributions that when superimposed qualitatively matches our measured loop, indicating MOKE in this configuration is only sensitive to the L1 layer.

**Magnetic Force Microscopy.** Further experiments and simulations were performed in order to examine the magnetisation configuration in the 3DNL. Fig 4a shows a magnetic force microscopy (MFM) image of the 3DNL surface after a saturating field along $H_1$. For clarity, the image has been processed to improve the contrast in the upper nanowires. Raw data is shown in Extended Data Fig 4. The upper-most layer of the lattice terminates in a series of bipod structures (L1 in Figs 2b and 2c). Strong contrast of the same complex pattern is measured in each bipod unit. Identical contrast is also observed after a saturating field in the plane of the substrate but perpendicular to $H_1$ (See Extended data Fig 7). In order to confirm the contrast is truly magnetic in origin and not an artefact of the topography, an image was measured in the same sample area but with a reversed tip magnetisation, as shown in Fig 4b. Here it is expected that positive colour contrast (yellow) should now be measured as negative colour contrast (black) and vice-versa. Close examination of individual bipod units in the two tip configurations (Fig 4c), show they have inverted contrast, confirming that the method is truly imaging the magnetic configuration of the 3DNL surface. Interestingly, contrast at the vertices (Fig 4d), where four wires meet, also exhibit an inversion of MFM contrast upon reversing the tip magnetisation. Raw data also shows this inversion of contrast as shown in Extended Data Fig 5.

**Discussion**

Figs 4e and 4f show simulated remanent states for a bipod structure whereby the individual nanowire magnetisation vectors were relaxed in the same and in opposing directions respectively. The red and yellow colour contrast represents the divergence of the magnetisation ($\nabla \cdot M$) which is an effective simulation of the MFM signal [48]. Comparison between Fig 4e and 4c/top shows good agreement, confirming the nanowires within the 3DNL are single domain. A more detailed vector plot of the magnetisation distribution within the bipod unit (Fig 4e Top) shows that the magnetisation smoothly rotates at the upper vertex area. Here the change in sign of $M_z$ yields MFM contrast at the vertex area as observed experimentally. Fig 4f shows a two-out state for the upper bipod unit. This high energy state consists of a vortex domain wall pinned on the vertex. When examining the hysteresis loop obtained from the lattice (Fig 3e), a single abrupt transition is obtained suggesting the wall nucleation field is sufficient to overcome any pinning at the vertex. This matches previous studies where it has been noted that in 2D connected ASI structures within the vortex regime, vertex pinning is not observed [21].

Magnetic nanowires of crescent-shaped cross-sectional geometry have not been considered extensively within the literature and certainly not in complex diamond lattice geometries. A key question here is the extent to which this nanowire geometry impacts the available remanent states when compared to a more idealised cylindrical shell geometry. In such a geometry, the vertex will have the idealised four-fold rotational symmetry lending itself to complete degeneracy of the ice-rule states (2-in/2-out states). One concern here is that the broken symmetry in our structures will yield vastly differently energies for ice-rule states. More symmetric structures are technologically very challenging and would require both the use of advanced non-linear optical techniques to obtain a spherical point spread function during the TPL process and a vapour-based deposition method to ensure all sidewalls are coated. Figs 4g-4m show the remanent state for tetrapod structures with all possible magnetisation configurations, except those that can be obtained by simple rotational transformations. To aid comparison with 2DASI literature, we have adopted the same vertex nomenclature. The vertex types where four-magnetisation vectors point in/out of a vertex were unstable and these instead relaxed into lower energy configurations. All simulations yield single domain nanowires indicating that the nanowires within our 3D nanolattice can approximate Ising spin behaviour allowing parallels to be drawn with the bulk frustrated materials known as spin-ice [5]. We emphasise that the nanowire configuration within our lattice maps directly onto the spin-ice lattice, allowing analogy with this bulk frustrated material. Contrary to what is seen in 2DASI where type 1 and type 2 vertices have different energy, computation of the energies (Extended Data Fig 6) in our structures shows that all type 1 and type 2 vertex energies agree to within 10%. This result is very exciting since it allows the possibility of realising a magnetic Coulomb phase, seen previously in layered systems [29]. However, in our system we note that replicating the geometry of the spin-ice lattice will yield a vanishing string tension between monopoles.

The simulated MFM contrast at the vertex within each of the configurations shown in Figs 4g-4m is distinct and depends upon the magnetisation within each of the underlying nanowires. Only one of the configurations (Fig 4i) shows adjacent lobes of inverted contrast at the vertex, matching the experimental signature shown in Fig 4d. Thus, upon reducing the saturating field along $H_1$ to remanence, the upper wires remain aligned with the field as expected, but 3D dipolar interactions within this frustrated geometry forces the lower wires to adopt the preferred low energy state. Overall, this yields one of the low energy ice-rule configurations whereby two-spins point into a vertex and two spins point out. Now that it has been demonstrated that 3D nanostructured frustrated lattices can be fabricated and that the spin texture can be measured via conventional magnetic microscopy, it opens up a huge array of potential experiments, analogous to those carried out upon 2D artificial spin-ice whereby the vertex type and correlations are measured in order to determine the rigidity of the ice-rule and long range ordering. Such experiments would require the development and rigorous testing of demagnetisation or thermalisation routines for 3D nanostructured samples but may ultimately provide deep insight into ground state ordering within bulk frustrated magnets.

In conclusion, we have demonstrated the fabrication and measurement of a 3D frustrated nanowire lattice. We have shown that the individual nanowires are single domain, magnetic switching of the lattice is dominated by the nucleation and propagation of vortex domain walls and this can be probed directly using optical magnetometry. Finite element simulations show that the low energy ice-rules states corresponding to type 1 and type 2 vertices in 2DASI are close in energy, suggesting our system is a candidate 3DASI material. Furthermore, we show directly using magnetic force microscopy experiments and finite element simulations that the vertex contrast can be used to elucidate the spin texture upon the lattice surface. Ice-rule violating defects within bulk spin-ice materials may be considered free monopoles in M and H. We note that the 3D symmetry of the lattice in our system will allow the realisation of analogous magnetic charges [6] that propagate freely and with the absence of any string tension seen in 2D artificial materials. The simplicity of our approach also lends itself to the realisation of a thermal 3DASI, by utilising thin NiFe layers. Finally, we note that our study will enable new avenues for the investigation of domain wall propagation through 3D magnetic nanowire lattices, allowing one to readily realise 3D magnetic racetrack systems as originally envisaged [32].

**Methods**

**Fabrication**

Glass coverslips (22 mm × 22 mm) with thickness varying between 0.16-0.19mm were used as substrates for this study. They were cleaned using acetone and isopropyl alcohol (IPA) solutions before immersion oil (Immersol 518 F) and a negative tone photoresist (IPL-780) were drop cast on to the lower and upper faces respectively. A diamond lattice covering a volume of 50 μm × 50 μm × 10 μm was then exposed within the photoresist using a two-photon lithography procedure. The samples were developed in propylene glycol monomethyl ether acetate for 20 minutes, removing any unexposed photoresist. Finally, the samples were cleaned in IPA and dried with a compressed air gun.

Permalloy ($Ni_{81}Fe_{19}$) of 99.99% purity was deposited via line of sight deposition. Samples were mounted at the top of a thermal evaporator. A 0.06 g ribbon of $Ni_{81}Fe_{19}$, cleaned in Isopropyl Alcohol, was placed in an alumina evaporation boat before evacuation of the chamber to a base pressure below $10^{-6}$ mBar. A uniform 50nm film was deposited at a rate of 0.2nm/s.

**Scanning electron microscopy (SEM) and energy dispersive x-ray analysis (EDX)**

Imaging was performed using a Zeiss Sigma HD Field Emission Gun Analytical SEM. Images were taken from top view as well as at a 45° tilt with respect to the substrate plane. Samples were mounted upon a SEM stub using carbon paint, which was placed in the SEM chamber. Prior to imaging, the chamber pressure was brought below $1 \times 10^{-4}$ mB. EDX compositional mapping was performed upon a sample using a 5 kV accelerating voltage for an extended period of time in order to maximise the resolution. Spectra from 10 locations were averaged to verify the intended 81:19 ratio of Ni:Fe in the sample.

**Magnetic Force Microscopy (MFM)**

MFM on the three-dimensional nanowire network was taken using a Bruker Dimension 3100 Atomic Force Microscope (AFM) operated in tapping mode. Prior to mounting, the commercial low moment MFM tips were magnetised parallel to the tip axis using a 0.5T permanent magnet. To magnetise the sample, a bespoke electromagnet was mounted on the surface of the AFM stage, allowing the application of a uniform magnetic field. The MFM tip was raised by several mm before application of the field. Images were taken, such that the fast scan axis was at an angle of 45 degrees to layer L1 of the lattice. MFM data was captured using a lift height of 100nm. MFM images were only taken once the feedback settings had been carefully optimised to ensure sample topography was being accurately measured.

**Magneto-optical Kerr Effect Magnetometry (MOKE)**

A 100 mW, 637 nm wavelength laser was attenuated to a power of approximately 6 mW, expanded to a diameter of 1 cm, and passed through a Glan–Taylor polarizer to obtain an s-polarized beam. The beam was then focused onto the sample using an achromatic doublet (f = 10 cm), to obtain a spot size of approximately 10 μm². The reflected beam was also collected using an achromatic doublet (f = 10 cm) and passed through a second Glan–Taylor polarizer, from which the transmitted and deflected beams were directed onto two amplified Si photodetectors, yielding the Kerr and reference signals, respectively. A variable neutral density filter was used to ensure that the reference and Kerr signals were of similar values. Subtraction of the reference from the Kerr signal compensates for any change in the laser intensity drift and also eliminates any small transverse Kerr effect from the signal.

**Finite element simulations**

A series of micro-magnetic simulations using finite element method discretisation were performed using the NMAG code [49]. These simulations are performed by numerical integration of the Landau-Liftshitz equation upon a finite element mesh. Typical $Ni_{81}Fe_{19}$ parameters were used, i.e. $M_s = 0.86 \times 10^6$ A m$^{-1}$, $A = 13 \times 10^{-12}$ J m$^{-1}$ with zero magnetocrystalline anisotropy. The simulations were performed at a temperature of 0 K that has previously been shown to capture the correct spin-texture seen in room temperature measurements but a systematic difference in coercivity (factor of ~5) is observed. The wire cross section is a crescent shape where the arcs subtend a 160° angle. The inner arc is defined from a circle with 80 nm radius corresponding to the 160 nm lateral feature size of the TPL system. Line of sight deposition results in a film thickness proportional to the scalar product of the deposition direction and the surface normal therefore, the outer arc is based on an ellipse with an 80 nm minor radius and 130 nm major radius. The length of the wires is set to 780nm, due to computational restraints and the wires are arranged as single wires, bipod and tetrapod structures. A comprehensive description of the geometries is given in the supplemental information. The geometries were meshed using adaptive mesh spacing with a lower limit of 3 nm and upper limit of 5 nm.

**Data Availability**

Information on the data underpinning the results presented here, including how to access them, can be found in the Cardiff University data catalogue.

**Acknowledgements**

SL acknowledges funding from the EPSRC (EP/R009147/1) as well as discussions with Prof Sean Giblin and Prof Wolfgang Langbein. We gratefully acknowledge technical support from Duncan Muir during EDX acquisition. We gratefully acknowledge support from Advanced Research Computing at Cardiff (ARCCA).

**Author Contributions**

SL conceived of the study and wrote the manuscript. AM carried out fabrication of 3D nanolattice samples, performed SEM, EDX, AFM and MFM upon the samples. MH carried out SEM and MOKE upon the samples. AV created the geometries and carried out finite element simulations upon single wire geometries, bipod geometries and tetrapod geometries and also performed SEM upon samples. AH carried out SEM and EDX analysis. All authors contributed to the final draft.

**Supplementary Figures**

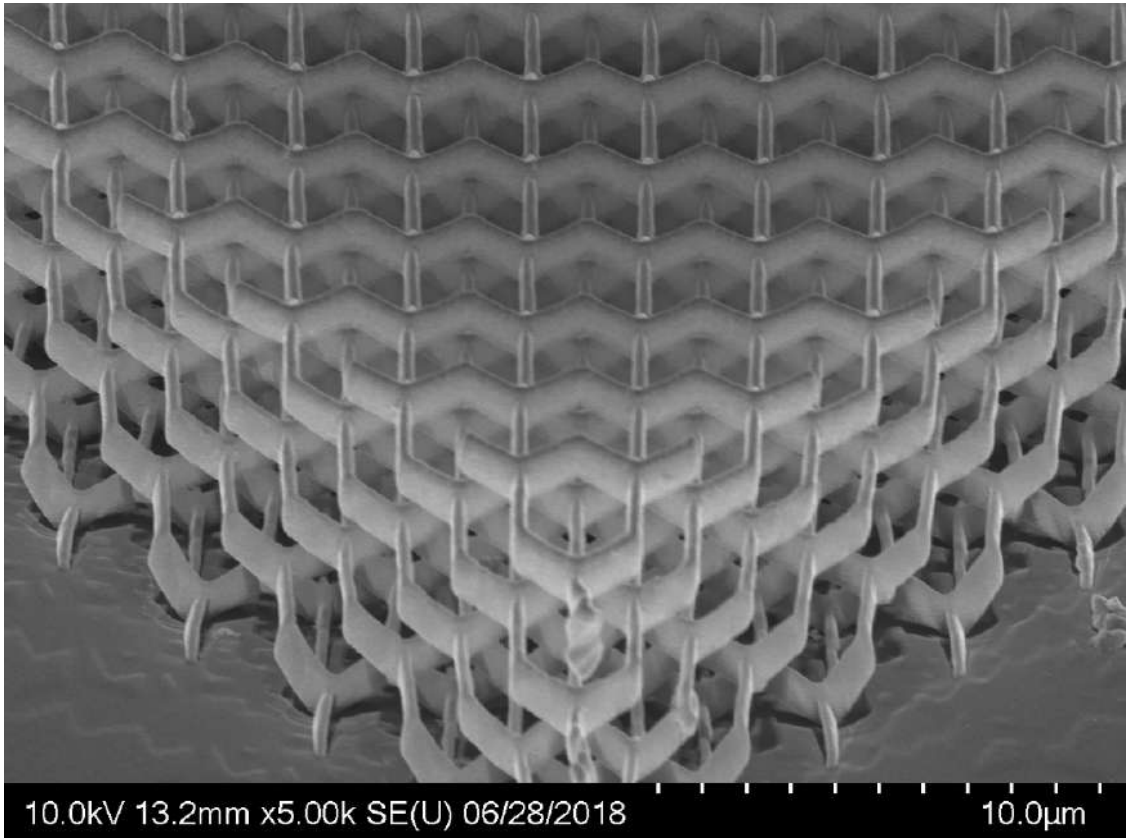

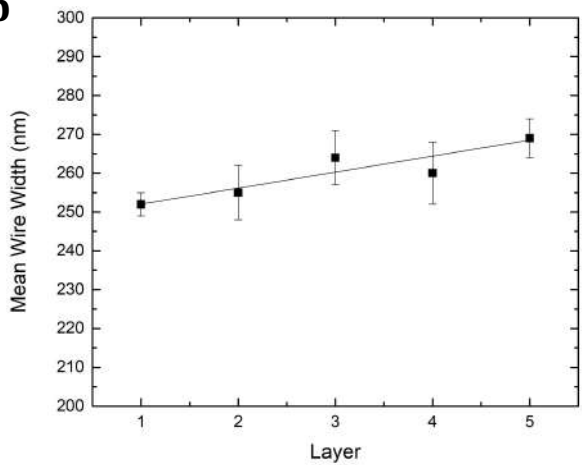

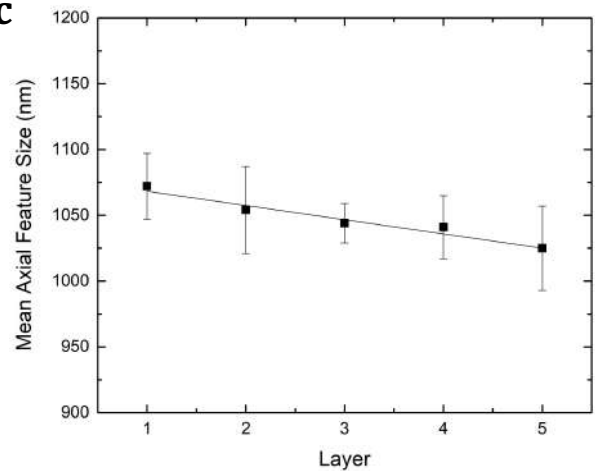

**Extended Data Figure 1: Variation of nanowire physical properties with lattice layer.** Note the sample was coated with approximately 4 x 10nm Au from each side to reduce charging during image acquisition. (a) SEM image of 3D lattice from 45-degree angle. Each of the individual layers are clearly visible. (b) Variation of nanowire lateral feature size with layer. (c) Variation of nanowire axial feature size with layer.

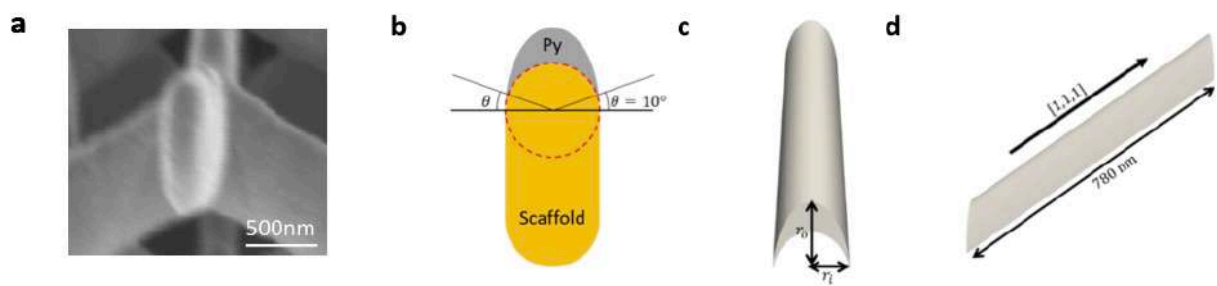

**Extended Data Figure 2: Design of 3D magnetic nanowires for finite element simulations.** (a) SEM image of polymer wire cross-section. Magnetic material is deposited onto the upper surface. (b) Diagram detailing how the simulated cross section was determined. (c) Diagram of a single nanowire showing the cross section. (d) Diagram of a single wire as seen from the side.

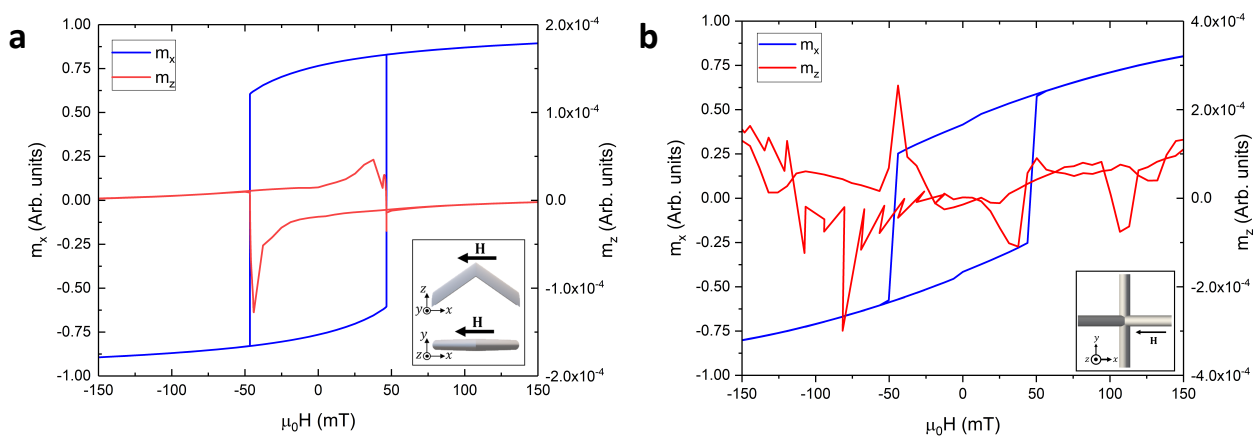

**Extended Data Figure 3: Simulated hysteresis loops upon (a) bipod and (b) tetrapod system.** For the bipod system the field was applied along the projection of the long axis as shown in inset. For the tetrapod system the field was applied along the projection of the upper wires as shown in inset.

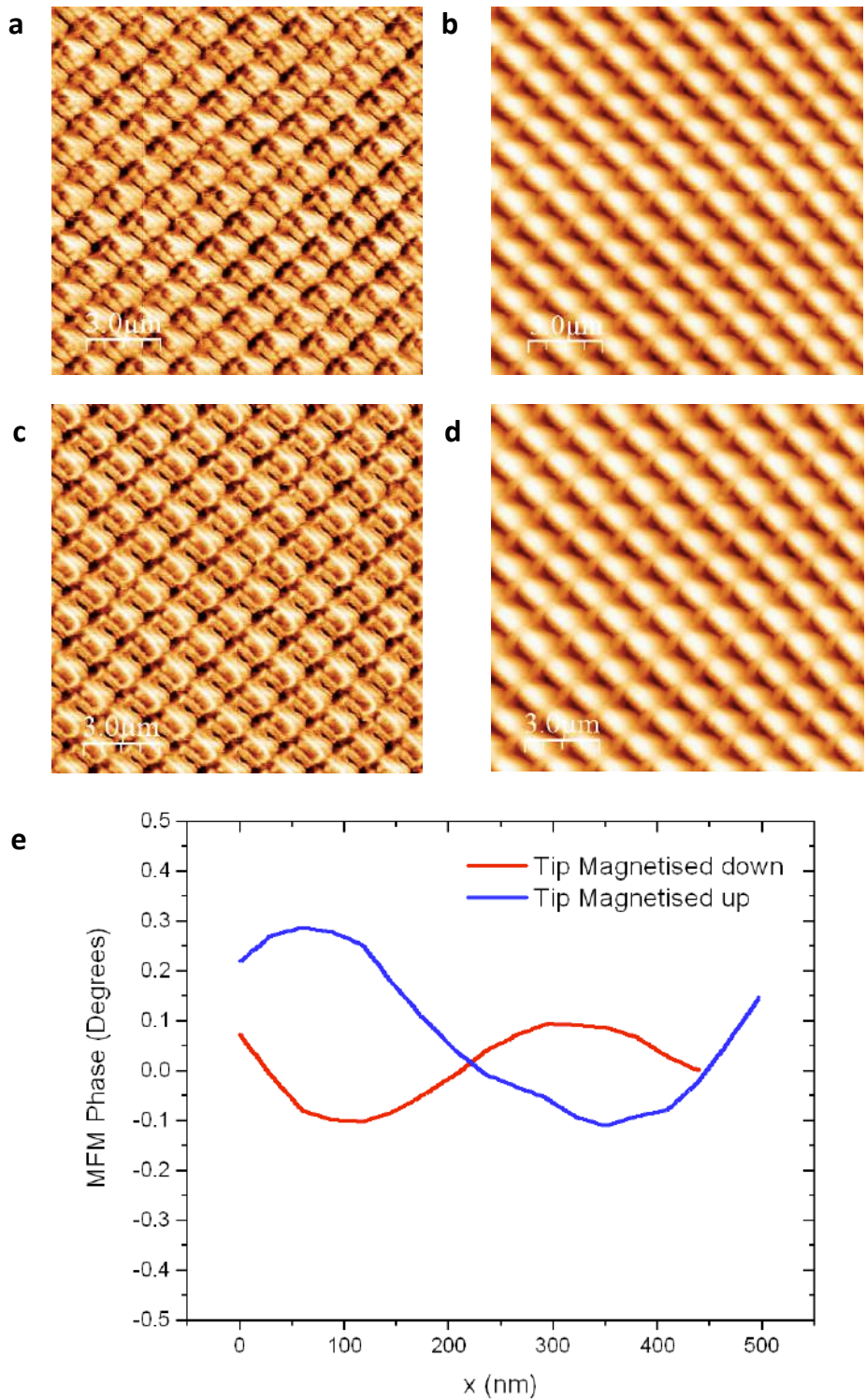

**Extended Data Figure 4:(a) MFM image of the 3D nanostructured lattice with tip magnetised down. (b) Corresponding AFM image. (c) MFM image of the 3D nanostructured crystal with tip magnetised up. (d) Corresponding AFM image. (e) Line scans across vertex where four nanowires meet. Each dataset has averaged 50 profiles from distinct vertices. The vertex contrast is seen to be inverted after tip reversal demonstrating this is magnetic contrast.**

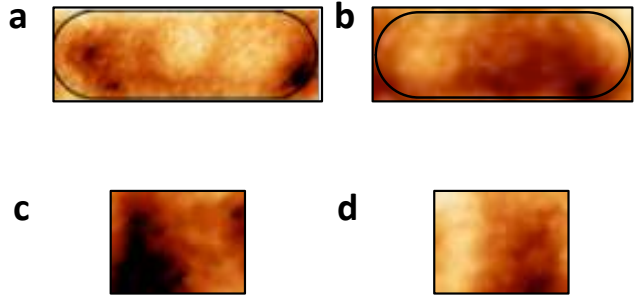

**Extended Data Figure 5:** (a) Raw MFM image of upper bipod unit with tip magnetised down. (b) Raw MFM image of upper bipod unit with tip magnetised up. (c) MFM image of vertex where four nanowires meet and with tip magnetised down. (d) MFM image of vertex where four nanowires meet and with tip magnetised up.

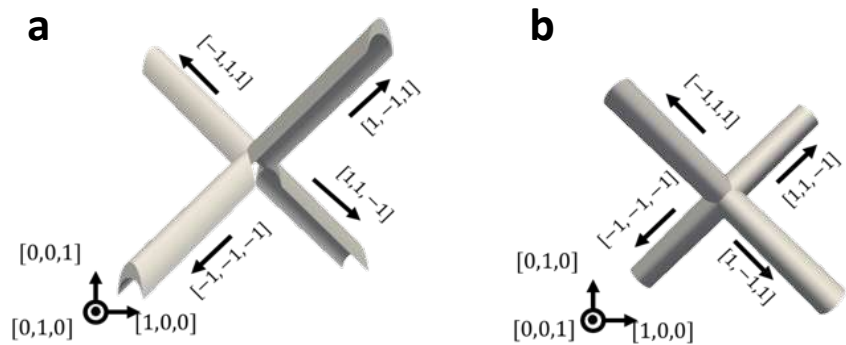

c

| ID | Vertex Type | Energy Density (J/m) |
|---|---|---|
| 0011 | Type I | 8342 |
| 1100 | Type I | 8342 |
| 0110 | Type II | 7745 |
| 1001 | Type II | 7746 |
| 1110 | Type III | 10447 |
| 1011 | Type III | 11683 |
| 0100 | Type III | 11276 |
| 0001 | Type III | 9907 |

**Extended Data Figure 6:** Total energy for magnetic tetrapod structures in different configurations. See supplementary information for explanation of ID.

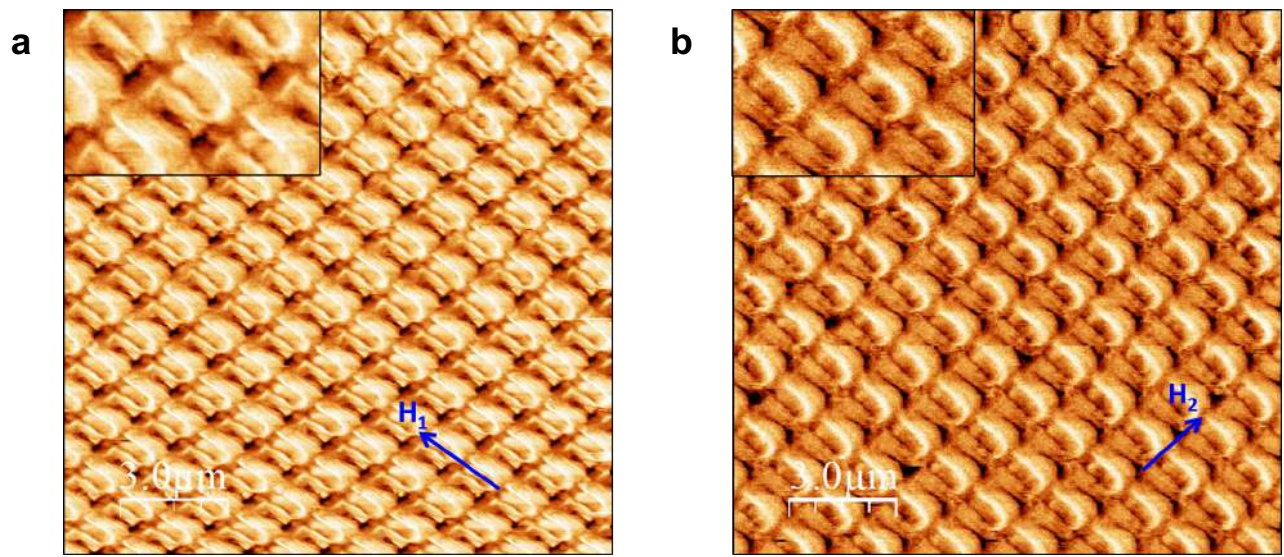

**Extended Data Figure 7: MFM images of the 3D nanostructured lattice taken after relaxing from saturation with an external magnetic field applied (a) parallel to the projection of L1 nanowires onto substrate ($H_1$) and (b) in the substrate plane but perpendicular to L1 ($H_2$).**

**Supplementary Discussion**

**Additional discussion upon 3DMOKE**

A key question when carrying out magneto-optical Kerr effect magnetometry (MOKE) upon a 3DNL is the degree of sensitivity of the signal to different sub-lattice layers in the system. One methodology to answer this is to compare the measured loop in Fig 3e to that of magnetic finite element simulations upon bipod and tetrapod units as shown in Extended Data Fig 3. Comparison of the remanence, as well as the overall loop shape allows one to infer the measured sub-lattices. Our measured MOKE loop is complex in shape, as detailed below, but we note it has a high remanence and square-like characteristic, more closely resembling the bipod characteristic shown in Fig 3e. Thus we conclude that with a 45° angle of incidence, the specular MOKE signal is mainly sensitive to sub-lattice L1 upon the 3DNL.

When studying 2D magnetic samples, the longitudinal or polar MOKE effect is chosen in order to obtain the information regarding the desired magnetisation component from the sample. In the 3D samples studied here, our nanowires make an angle of 35° with respect to the substrate meaning that an angle of incidence of 45° will yield both a longitudinal and polar MOKE signal. Thus the loop shown in Fig 3e is showing a superposition of these two signals.

The longitudinal and polar features within the loop can be determined by comparison to finite element simulations upon bipod structures. For the longitudinal loop, we expect a maximum positive signal at positive saturation and an abrupt switching at the coercive field of the nanowires after which the magnetisation starts to cant into the plane of the sample, providing a maximum negative longitudinal signal at negative saturation. For the polar loop, at saturation there is no net signal since the nanowire magnetisations have zero $m_z$ component. Upon reducing the in-plane field, the strong demagnetisation field within the nanowires starts to take effect and the magnetisation starts to rotate into the direction of the nanowire long axis yielding a polar component. This increases as the field decreases from saturation but is not seen over our experimental field range. At the coercive field of the nanowires, the polar component changes abruptly. The exact form taken here depends upon the reversal of the individual wires within bipod units. Assuming domain walls nucleate from the bipod extremities, even a very small difference in the domain wall nucleation fields will lead to partial cancellation of the polar component and a net polar signal is seen as shown by the green arrow in Fig 3e. Micro-magnetic simulations of bipod structures shown in Fig 3f confirm these arguments. Here it can be seen that a linear combination of $m_x$ (longitudinal) and $m_z$ (polar), reproduces the qualitative form of our measured MOKE loop.

**Additional discussion upon magnetic force microscopy data**

A raw magnetic force microscopy (MFM) image with the tip magnetised down and after a saturating field along $H_1$ (projection of upper nanowires) is shown in Extended Data Fig 4a. The image is seen to be complex with contrast originating from both the nanowires and space within the voids. Since we are using a two-pass technique (See methods), the MFM signal can only be assumed to be accurate in areas where well-defined topography is measured. The accurate information can be found by comparing the MFM image in Extended Data Fig 4a with its topography image shown in Extended Data Fig 4b. As discussed in the manuscript, it can be seen that the contrast upon the upper nanowires invert upon reversing the tip magnetisation as can be seen by comparing contrast in Extended Data Fig 4a and Extended Data Fig 4c.

In order to examine only the contrast associated with the upper-most wires, the raw MFM images were multiplied by the topography image shown in Extended Data Fig 4b. This removes contrast associated within the void areas but maintains contrast associated with well-defined topography in upper wires and at vertices where four wires meet. The resulting images were rotated such that the upper wire projection were along the image horizontal and increased in magnification to improve visibility of magnetic states. Analysis of the raw images also shows clear inversion of MFM contrast upon switching tip magnetisation as shown in Extended Data Fig 5. To confirm this is indeed the case, 50 line profiles were taken from individual vertices in images obtained with tip magnetised up and down. The result of this is shown in Extended Data Fig 4e and the signal is clearly seen to invert demonstrating the vertex contrast is magnetic in origin.

**Supplementary Methods: Finite element modelling**

From SEM images, the lateral feature size is found to be 160 nm and the cross-sectional shape of a voxel in our samples is found to be a rounded rectangle upon which $Ni_{81}Fe_{19}$ is deposited as shown in Extended Data Fig 2a. As $Ni_{81}Fe_{19}$ is deposited using evaporation, material is only present on the upper part of the polymer scaffold and the inner arc of the magnetic material may then be approximated as a semicircle with an 80 nm radius. From geometric arguments supported by SEM images and the literature [1,2], the film thickness is proportional to the scalar product of the deposition direction and the normal of the surface giving a maximum thickness of 50 nm at the apex. The $Ni_{81}Fe_{19}$ material is not expected to subtend the full arc of the semicircle and the geometry is clipped at 80° angles from the apex (Extended Data Fig 2b). The final nanowire design is shown in Extended Data Figs 2c and 2d.

In order to determine the stability of different vertex types seen in 2DASI, type 1, type 2, type 3 and type 4 vertices were relaxed in the absence of an applied field. In addition, the energy of each vertex type was extracted from the finite element simulations. The energies can be found in Extended Data Fig 6. Here the ID is a 4-bit binary number where each bit corresponds to one of the wires in the tetrapod and a 1 indicates that the wire is magnetised towards the vertex and a 0 indicates the wire magnetisation away from the vertex.

Starting from the left most bit, each bit refers to:

Lower wire along [1,1,-1]

Lower wire along [-1,-1,-1]

Upper wire along [-1,1,1]

Upper wire along [1,-1,1]

In order to understand the experimental magnetometry upon the 3D nanowire lattice, simulated hysteresis loops were obtained for nanowires placed into a bipod or tetrapod geometry as shown in Extended Data Fig 3.

**Supplementary References**

[1] Albrecht, M., et al., Magnetic multilayers on nanospheres. Nature Materials, 2005. 4(3): p. 203-206.

[2] Streubel, R., et al., Magnetically Capped Rolled-up Nanomembranes. Nano Letters, 2012. 12(8): p. 3961-3966.